\documentclass[a4paper,11pt]{article}
\usepackage{pos}
\usepackage{lineno}  

\title{EPOS4 Model Predictions for Global Observables in Pb–Pb Collisions at $\sqrt{s_{NN}}$ = 5.36 TeV}

\author*[a]{Hirak Kumar Koley}
\author[a]{Subikash Choudhury}
\author[a, b]{Mitali Mondal}

\affiliation[a]{Nuclear and Particle Physics Research Centre, Department of Physics, Jadavpur University,\\ Kolkata - 700032, India}

\affiliation[b]{School of Studies in Environmental Radiation and Archaeological Sciences, Jadavpur University,\\ Kolkata - 700032, India}

\emailAdd{hirak.koley@gmail.com}

\abstract{

The study of the Quark-Gluon Plasma (QGP), a deconfined state of nuclear matter, remains a central focus of high-energy heavy-ion collision experiments. The recent operation of the Large Hadron Collider (LHC) in Run 3 at the new center-of-mass energy of $\sqrt{s_{NN}}=5.36$ TeV necessitates theoretical predictions to characterize the energy dependence and bulk properties of the medium. In this study, we present comprehensive EPOS4 model predictions for key global observables in Pb-Pb collisions at $5.36$ TeV. We focus on the centrality dependence of the charged-particle pseudorapidity density ($dN_{ch}/d\eta$), integrated yields ($dN/dy$), mean transverse momentum ($\langle p_{T}\rangle$) for light-flavor hadrons ($\pi^{\pm}, K^{\pm}, p(\bar{p})$), and the charged particle nuclear modification factor ($R_{AA}$). The EPOS4 framework successfully captures the strong mass-dependent rise of $\langle p_{T}\rangle$ with multiplicity, a definitive signature of collective radial flow. Furthermore, the predicted charged hadron $R_{AA}$ demonstrates a clear suppression, consistent with energy loss mechanisms incorporated into the model. By comparing these predictions to existing $5.02$ TeV data, we demonstrate that the EPOS4 model offers a consistent and robust description of heavy-ion dynamics, projecting minimal energy evolution for these bulk and hard-probe observables between the two energies.

}

\FullConference{The European Physical Society Conference on High Energy Physics (EPS-HEP2025)\\
7-11 July 2025\\
Marseille, France\\}

\begin{document}


\maketitle

\section{Introduction}

The Quark-Gluon Plasma (QGP) \cite{ref_1}, a transient, deconfined state of quarks and gluons, is the defining subject of high-energy heavy-ion collision physics. The LHC has provided a wealth of data in Run 3, yet characterizing the QGP's properties at varying collision energies is essential to constrain theoretical models. With the LHC now operating at a new collision energy of $\sqrt{s_{NN}}=5.36$ TeV for heavy-ions, robust theoretical predictions are vital to validate the hydrodynamic and transport mechanisms implemented in state-of-the-art event generators.
Key global observables act as essential probes of the system's evolution. The charged particle pseudorapidity density ($\mathrm{d}N_{\mathrm{ch}}/\mathrm{d}\eta$) provides critical information regarding the initial conditions and stopping power, reflecting the energy density achieved. At the differential level, the transverse momentum ($p_{T}$) spectra and integrated particle yields of identified light-flavor hadrons probe the interplay between particle production and collective dynamics. The characteristic mass hierarchy observed in the mean transverse momentum ($\langle p_{T}\rangle$) serves as the strongest experimental signature for radial flow—the collective outward expansion of the dense medium. Furthermore, at high $p_{T}$, the nuclear modification factor ($R_{AA}$) quantifies the suppression of hadron production, serving as the definitive measure of jet quenching and QGP opacity.
In this contribution, we present detailed predictions for these global and differential observables in Pb-Pb collisions at $\sqrt{s_{NN}}=5.36$ TeV, utilizing the EPOS4 model \cite{EPOS4}. Our analysis covers the centrality dependence of $\mathrm{d}N_{\mathrm{ch}}/\mathrm{d}\eta$, $\langle p_{T}\rangle$, and $\mathrm{d}N/\mathrm{d}y$ for $\pi^{\pm}$, $K^{\pm}$, and $(p,\bar{p})$. Crucially, the $5.36$ TeV predictions are systematically compared against available experimental measurements from the ALICE and CMS collaborations at the nearby Run 2 energy of $\sqrt{s_{NN}}=5.02$ TeV [3-6]. This systematic comparison is essential to validate EPOS4's predictive power, explore energy dependence, and ensure consistency with known LHC dynamics.

\section{Results and Discussions}
\subsection{Charged Particle Pseudorapidity and Multiplicity Density Distributions}

\begin{figure}[h]
\centering
\includegraphics[width=0.32\textwidth]{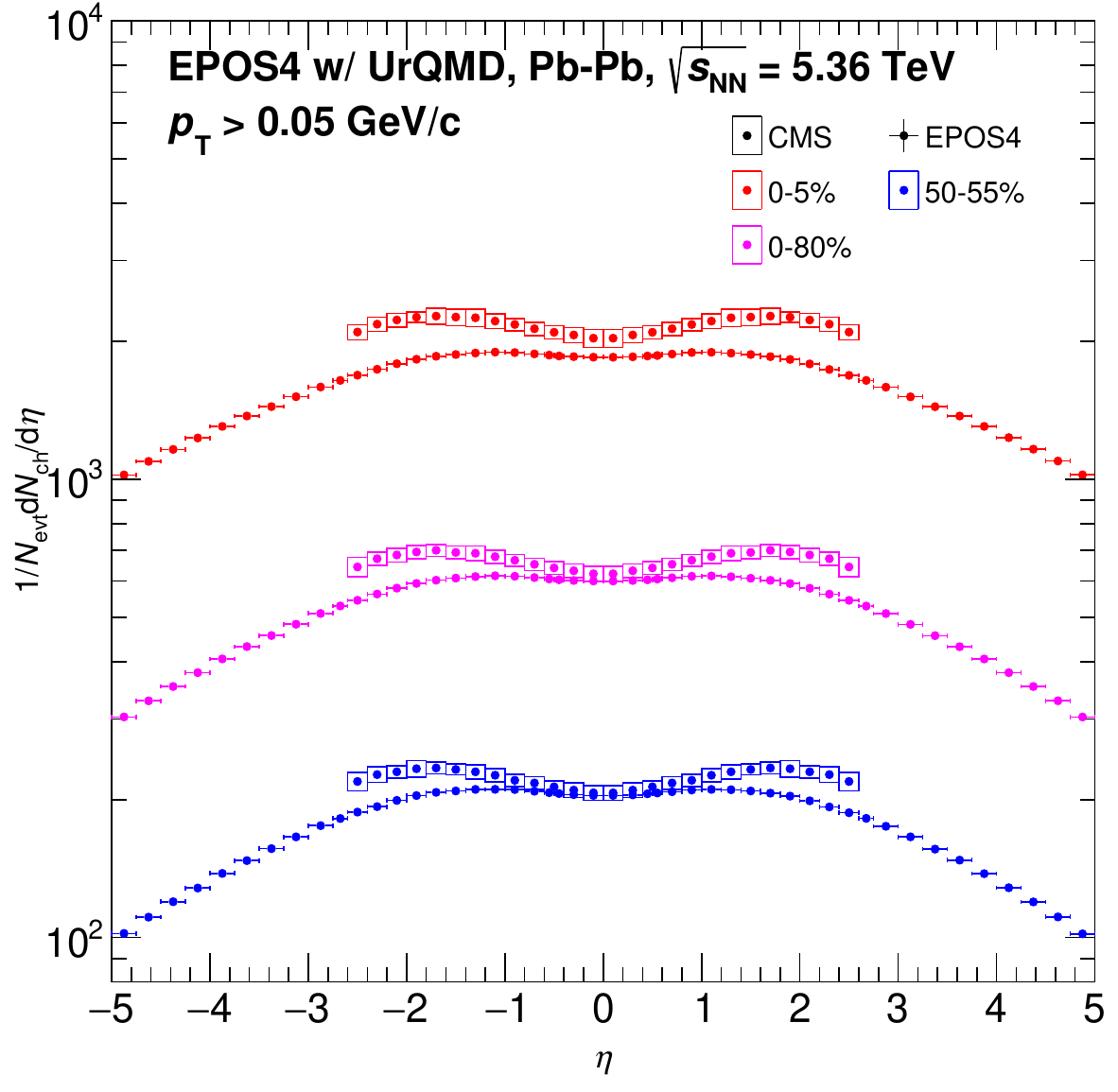}
\includegraphics[width=0.32\textwidth]{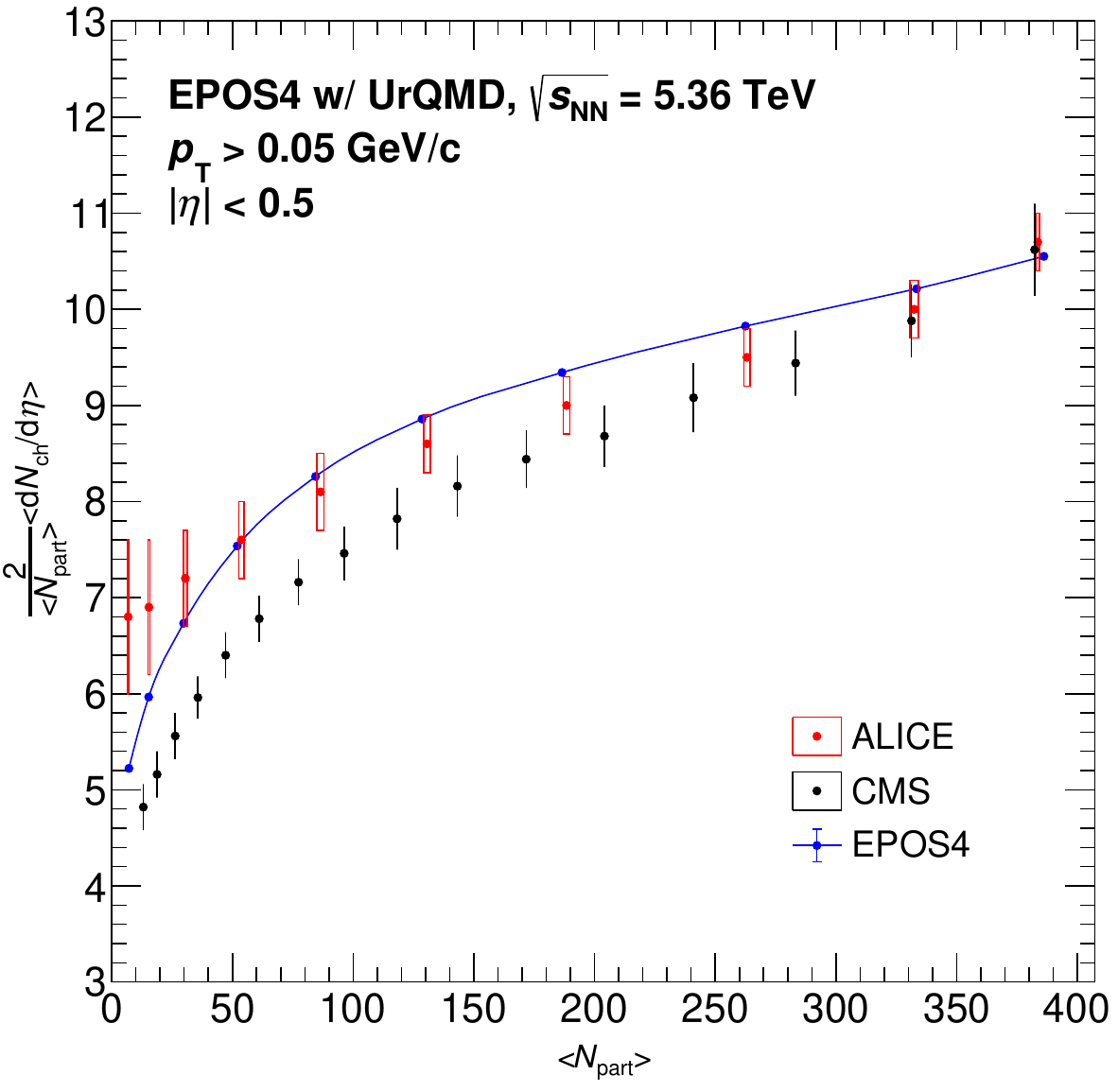}
\caption{EPOS4 predictions for (left) the charged particle pseudorapidity density for different centrality classes and (right) the charged particle multiplicity density per participant pair $\frac{2}{\langle N_{part} \rangle} \langle dN_{\text{ch}}/d\eta \rangle$ at mid-rapidity ($|\eta|$<0.5) as a function of $\langle N_{part} \rangle$, compared to ALICE \cite{ALICE_mult_ref} and CMS \cite{CMS_etadist_ref} measurements in Pb-Pb collisions at $\sqrt{s_{\mathrm{NN}}}$ = 5.36 TeV.}
\label{fig:etamultdist}
\end{figure}

 The left panel of Fig. \ref{fig:etamultdist} presents the EPOS4 predictions for the $dN_{\text{ch}}/d\eta$ distributions in Pb-Pb collisions at $\sqrt{s_{\mathrm{NN}}}$ = 5.36 TeV for three different centrality classes: 0-5$\%$ (most central), 50-55$\%$, and the integrated 0-80$\%$ centrality. The distributions are calculated for charged particles with $p_{\mathrm{T}}$ > 0.05 GeV/c.
The EPOS4 model, which incorporates the effect of final state interactions (labeled "w/ UrQMD"), predicts a broad, symmetric distribution in pseudorapidity ($\eta$), characteristic of high-energy nuclear collisions where the center-of-mass frame is the same as the detector frame. As expected, the density of produced charged particles is highest in the most central collisions (0-5$\%$), demonstrating a clear correlation between particle production and the degree of nuclear overlap. The distribution also exhibits a characteristic dip near mid-rapidity ($\eta\approx$ 0). The full $\eta$ coverage of the EPOS4 calculation allows for a comprehensive study of nuclear stopping and particle production over the entire kinematic range.

Further investigation into the bulk production is performed by examining the charged particle multiplicity density per pair of participating nucleons, $\frac{2}{\langle N_{part} \rangle} \langle dN_{\text{ch}}/d\eta \rangle$, at mid-rapidity ($|\eta|$<0.5). This quantity characterizes the normalized particle yield as a function of collision centrality, effectively probing the transition from hard-scattering dominated production to soft, bulk production.
The right panel of Fig. \ref{fig:etamultdist} shows the EPOS4 predictions for this normalized multiplicity density as a function of the average number of participating nucleons, $\langle N_{part} \rangle$. 
The predictions demonstrate a clear increase with $\langle N_{part} \rangle$, indicating that particle production scales faster than linearly with the number of participating nucleons. Within the EPOS4 framework, this behaviour originates from the increasing core fraction and the resulting enhancement of collective, hydrodynamic particle production in central collisions.
We compare these predictions to existing experimental data from the ALICE and CMS collaboration at $\sqrt{s_{\mathrm{NN}}}$ = 5.36 TeV \cite{ALICE_mult_ref, CMS_etadist_ref}, where the EPOS4 calculations show a consistent trend with the experimental observations. 
The deviation observed between the EPOS4 curve and the CMS data, particularly from peripheral to mid-central collisions, contrasts with the good agreement seen with the ALICE measurements. This difference might be due to variations in measurement techniques or event selection.

\subsection{Spectra, Yields and \texorpdfstring{$\langle p_{T} \rangle$}{<pT>} of Light Flavored Hadrons}

\begin{figure}[h]
\centering
\includegraphics[width=0.32\textwidth]{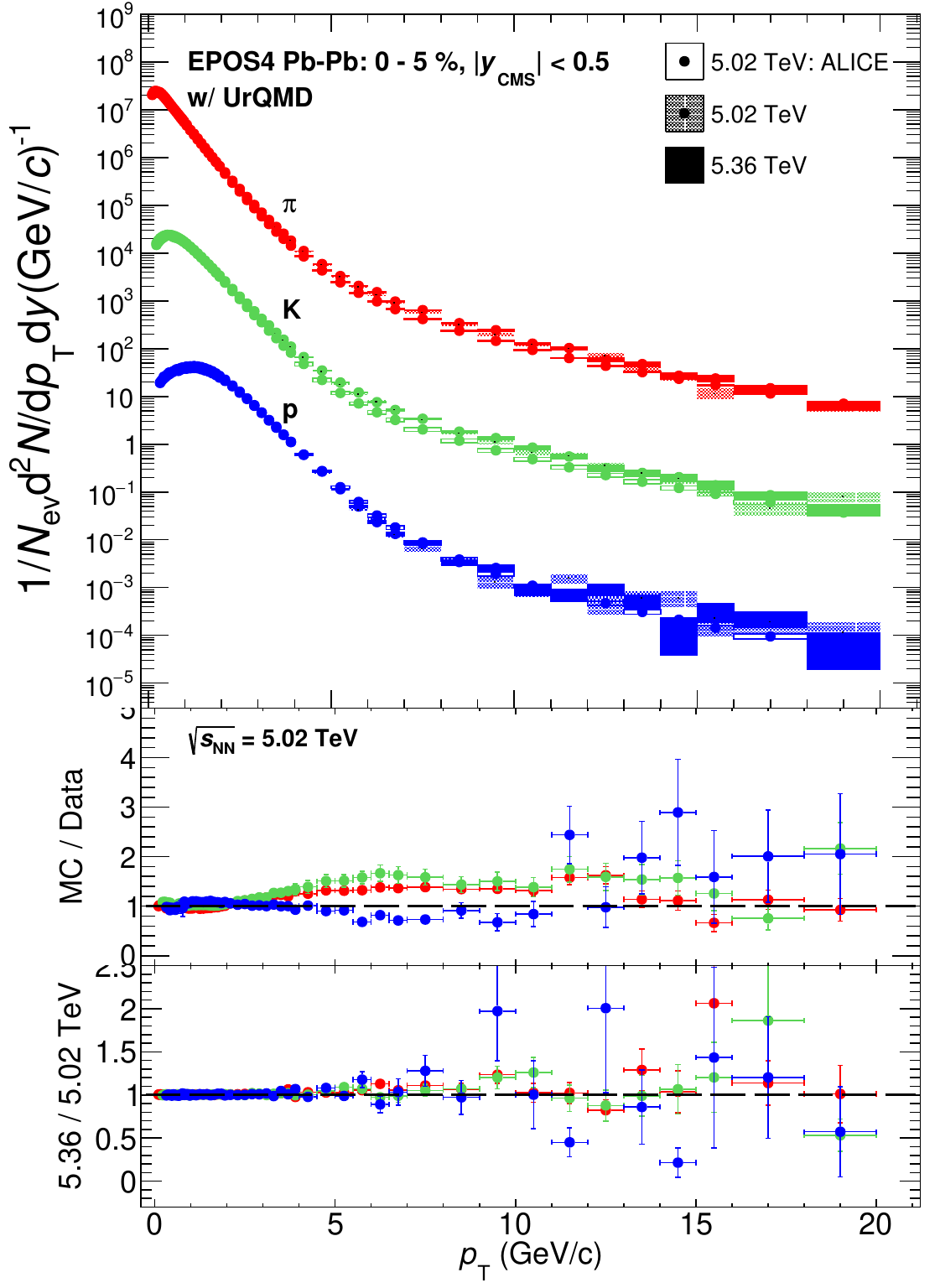}
\includegraphics[width=0.32\textwidth]{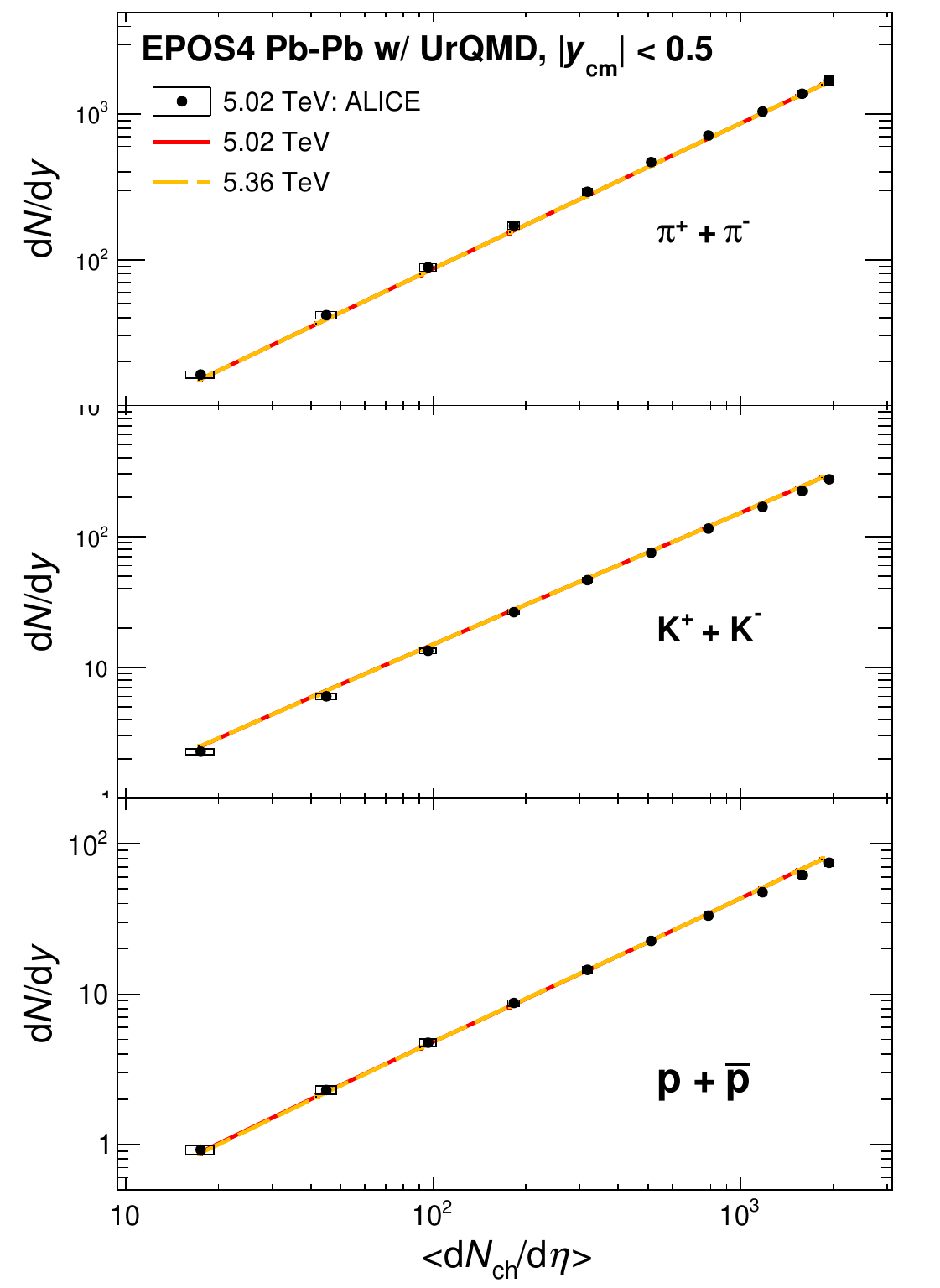}
\includegraphics[width=0.32\textwidth]{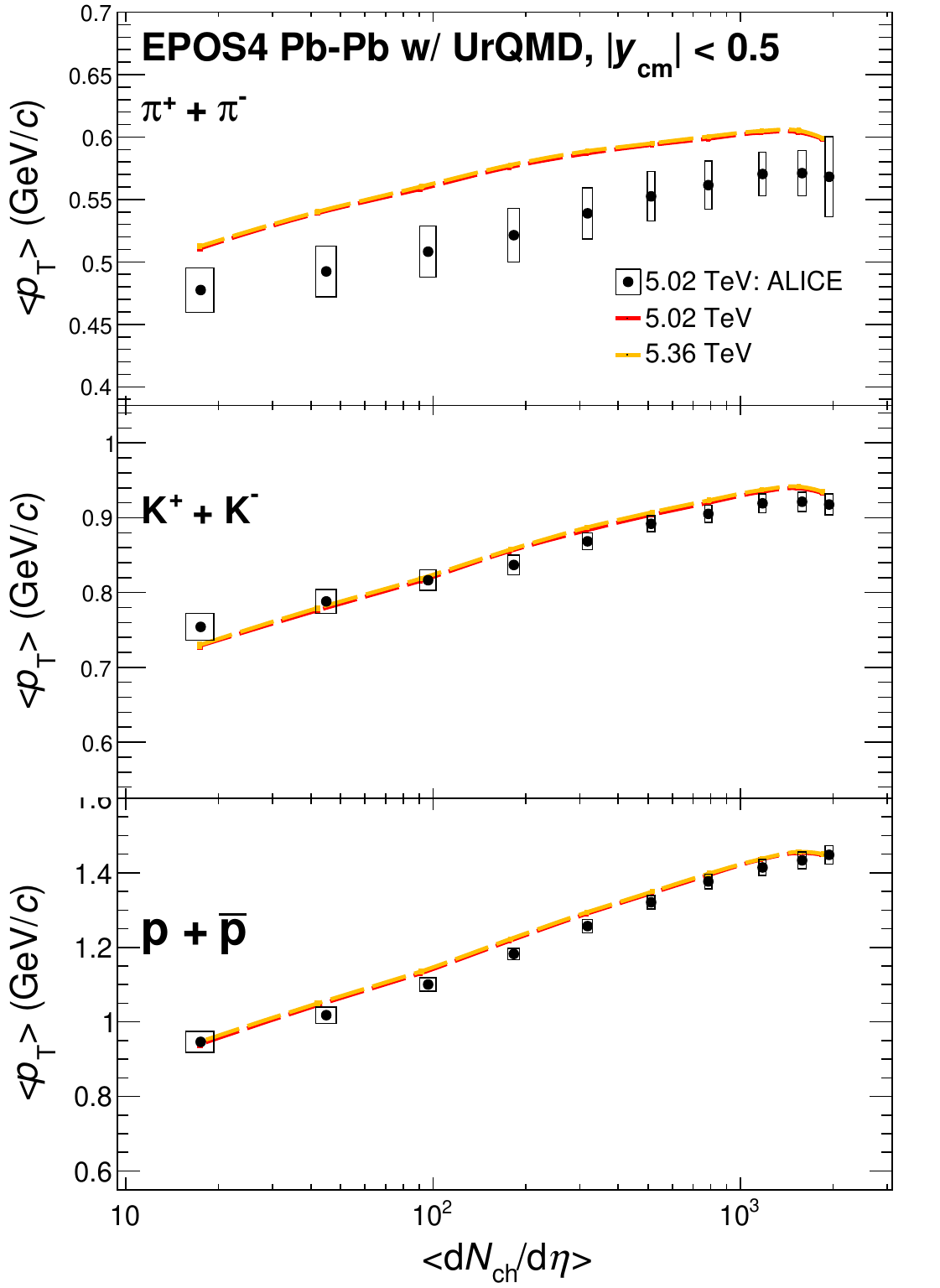}
\caption{EPOS4 predictions for (left) $p_{\mathrm{T}}$ spectra, (middle) integrated yields (dN/dy), and (right) mean $p_{\mathrm{T}}$ for $\pi^{\pm}$, $K^{\pm}$, and $p, \overline{p}$ at $|y_{cm}|$<0.5 in Pb-Pb collisions, compared with ALICE data \cite{ALICE_ptspec_ref}.}
\label{fig:ptdist}
\end{figure}

 The differential yield of identified light-flavor hadrons provides insights into the thermodynamic properties and collective expansion of the QGP. 
 The left panel of Fig. \ref{fig:ptdist} presents the EPOS4-predicted $p_{\mathrm{T}}$ spectra for $\pi^{\pm}$, $K^{\pm}$, and $p, \overline{p}$ at mid-rapidity ($|y_{cm}|$<0.5) for the most central (0-5$\%$) Pb-Pb collisions. 
 The spectra exhibit a characteristic exponential decay at low $p_{\mathrm{T}}$ followed by a power-law tail at high $p_{\mathrm{T}}$. 
 A clear mass hierarchy is visible at low $p_{\mathrm{T}}$, with heavier particles having steeper slopes, which is consistent with the radial flow generated by the collective expansion of the medium.
The results also include a comparison between the EPOS4 predictions at the two energies: 5.36 TeV and 5.02 TeV. The overall shape and magnitude of the 5.02 TeV spectrum are consistent with ALICE measurements \cite{ALICE_ptspec_ref}. The lower panel displays the ratio of the EPOS4 5.36 TeV prediction to the 5.02 TeV prediction, demonstrating a minimal change in the spectral shape, with an expected uniform enhancement in yield reflecting the increase in collision energy.

To characterize the system's thermodynamic evolution, we study the centrality dependence of the integrated yields (dN/dy) and the mean $p_{\mathrm{T}}$.
The middle panel of Fig. \ref{fig:ptdist} illustrates the integrated yield of $\pi^{\pm}$, $K^{\pm}$, and $p, \overline{p}$ as a function of the charged particle multiplicity density, $\langle dN_{\text{ch}}/d\eta \rangle$. For all three hadron species, the yields increase sharply with $\langle dN_{\text{ch}}/d\eta \rangle$, spanning almost two orders of magnitude from peripheral to central collisions. This strong correlation is expected, as higher multiplicity is directly linked to higher energy density. The EPOS4 predictions at 5.36 TeV and 5.02 TeV show excellent agreement with the ALICE data at 5.02 TeV over the entire range, and the energy dependence between the two EPOS4 predictions is negligible for the light-flavor hadron yields.
Finally, the right panel of Fig. \ref{fig:ptdist} shows $\langle p_{\mathrm{T}} \rangle$ for the three identified hadron species as a function of $\langle dN_{\text{ch}}/d\eta \rangle$. The $\langle p_{\mathrm{T}} \rangle$ exhibits a characteristic monotonic increase with multiplicity for all species. Crucially, the increase is mass-dependent, with $p, \overline{p}$ showing the steepest rise, followed by $K^{\pm}$, and then $\pi^{\pm}$. This distinct mass hierarchy in $\langle p_{\mathrm{T}} \rangle$ is a hallmark of strong radial flow developed in the dense QGP medium, where the collective expansion velocity is transferred to the individual particles. Both the 5.36 TeV and 5.02 TeV EPOS4 predictions accurately capture this trend and agree well with the ALICE measurements across the centrality range. However, the model slightly overestimates the $\langle p_{\mathrm{T}} \rangle$ of pions, suggesting that the soft particle production or freeze-out dynamics in EPOS4 may require further tuning to fully capture the light-hadron sector.

\subsection{Charged particle Nuclear Modification Factor}

The nuclear modification factor, $R_{AA}$, is the definitive measure of the energy loss experienced by hard-scattered partons traversing the QGP. It is defined as:
$$ R_{AA}(p_T) = \frac{d^2N^{AA}/dp_Tdy}{\langle N_{\text{coll}}\rangle d^2N^{pp}/dp_Tdy} $$
A value of $R_{AA}$ < 1 signifies hadron yield suppression relative to binary scaling, which is the primary evidence for jet quenching in the hot and dense QGP medium.
 \begin{figure}[h]
\centering
\includegraphics[width=0.32\textwidth]{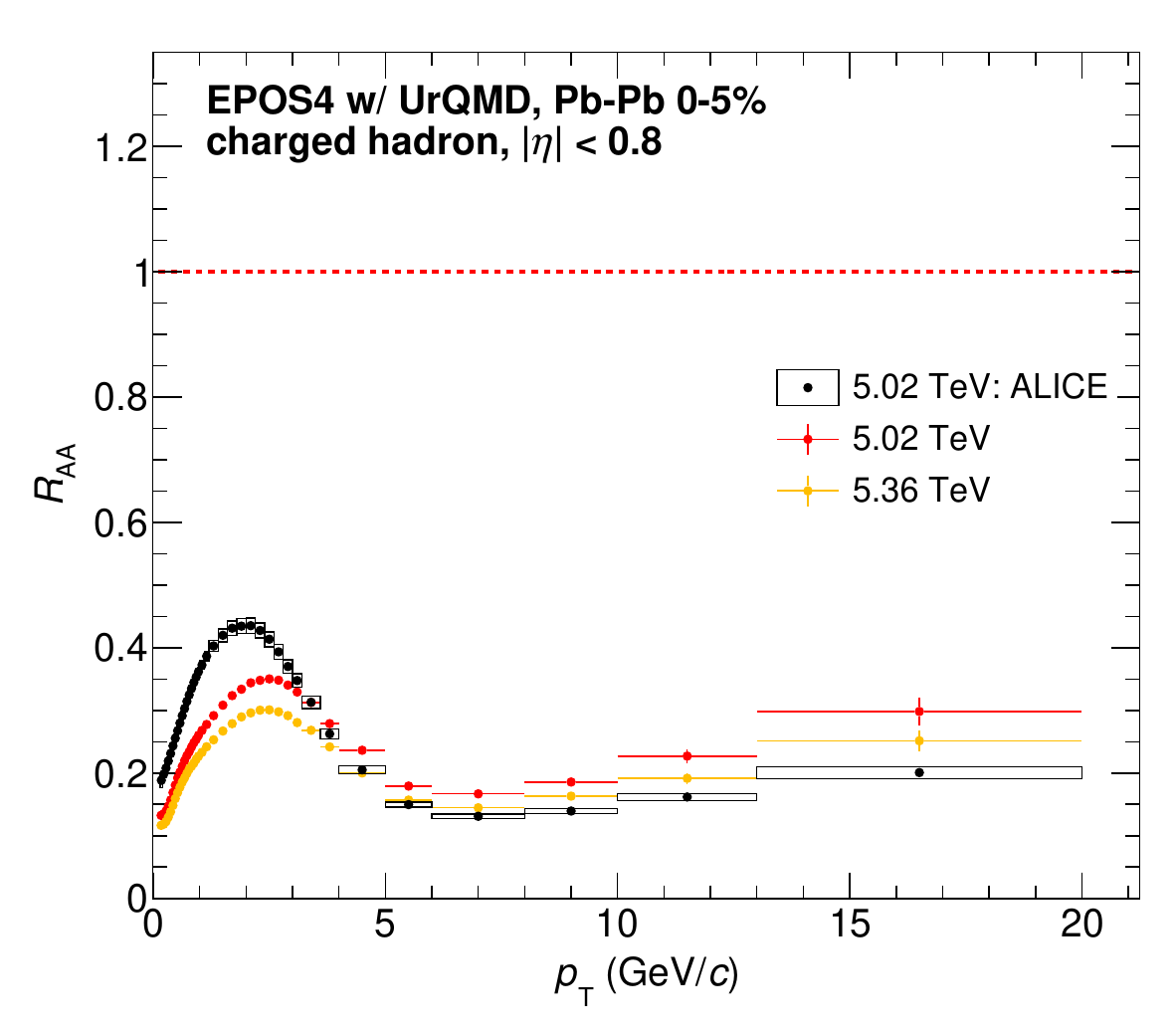}
\caption{EPOS4 predictions for charged hadron nuclear modification factor ($R_{AA}$) in 0-5$\%$ central Pb-Pb collisions at $\sqrt{s_{\mathrm{NN}}}$ = 5.02 TeV and $|\eta|$<0.8, compared with ALICE \cite{ALICE_Raa_ref}.}
\label{fig:Raa}
\end{figure}
In Figure \ref{fig:Raa}, we present the EPOS4 model prediction for the $R_{AA}$ of charged hadrons in 0-5$\%$ central Pb-Pb collisions at $\sqrt{s_{\mathrm{NN}}}$ =5.02 and 5.36 TeV, compared with ALICE Run 2 measurements at 5.02 TeV.
The EPOS4 calculation accurately captures the overall trend observed in the data. The $R_{AA}$ is suppressed across the entire $p_{\mathrm{T}}$ range shown, falling significantly below the unity baseline. We observe a minimum suppression around $p_{\mathrm{T}}\approx$ 2 GeV/c, which is primarily influenced by the contribution of soft, flowing particles.
At higher transverse momenta ($p_{\mathrm{T}} > 6$ GeV/$c$), the $R_{AA}$ plateaus around 0.2, indicating strong suppression of high-$p_{\mathrm{T}}$ particle production. This suppression reflects substantial energy loss of partons traversing the dense medium, validating the hydrodynamic and energy-loss mechanisms in EPOS4. The predicted $R_{AA}$ at 5.36 TeV closely follows the 5.02 TeV trend, confirming the stability of the suppression pattern with increasing collision energy. Nevertheless, EPOS4 slightly underestimates the data in the intermediate-$p_{\mathrm{T}}$ region, suggesting that refinements to the energy-loss parameters or hadronization modeling could improve agreement.

\section{Summary}

We presented EPOS4 model predictions for key global and differential observables in Pb-Pb collisions at $\sqrt{s_{\mathrm{NN}}}$ = 5.36 TeV, providing critical input for future analysis at the LHC. The calculations were validated against available experimental data at $\sqrt{s_{\mathrm{NN}}}$ = 5.02 TeV.
Regarding the bulk properties, the EPOS4 model successfully describes the centrality dependence of the charged particle multiplicity density, ($\frac{2}{\langle N_{part} \rangle} \langle dN_{\text{ch}}/d\eta \rangle$), and accurately predicts the overall shape of the pseudorapidity distribution, $dN_{\text{ch}}/d\eta$. The predictions for the identified light-flavor hadrons ($\pi^{\pm}$, $K^{\pm}$, and $p, \overline{p}$) reinforce the picture of a collectively expanding medium, characterized by a smooth, strong increase in the integrated yields with event multiplicity, and a pronounced, mass-dependent rise in the mean $p_{\mathrm{T}}$ as a function of multiplicity, serving as the strongest evidence for the development of significant radial flow.
Finally, the predicted nuclear modification factor ($R_{AA}$) for charged hadrons demonstrates strong suppression ($R_{AA}\approx$ 0.2) at high $p_{\mathrm{T}}$ in central collisions, confirming the essential role of jet quenching in the model's description of hard probes. Overall, the EPOS4 framework provides a consistent and robust description of heavy-ion collision dynamics at 5.02 TeV and offers reliable predictions for the next generation of measurements at 5.36 TeV, suggesting only minimal energy evolution for these observables between the two energies.

\begin{acknowledgments}
The authors gratefully acknowledges financial support from the Department of Science and Technology, Government of India, under the “Mega Facilities in Basic Science Research” scheme.
\end{acknowledgments}

\end{document}